# Research Power Ranking:

Adapting the Elo System to Quantify Scientist Evaluation

Eldar Knar[1]

Tengrion, Astana, Republic of Kazakhstan

eldarknar@gmail.com
https://orcid.org/0000-0002-7490-8375

**Abstract**

This paper presents an original model for assessing scientific productivity, research power ranking (RPR), which is based on the adaptation of the Elo rating system to the context of scientific activity.

Unlike traditional scientometric indicators, RPR accounts for the dynamics, multidimensionality, and relativity of research power.

The model comprises three components—fundamental, applied, and commercial activity—each represented by a separate rating and updated on the basis of probabilistic "scientific games" analogous to chess matches.

The scientific rating of each researcher is calculated as a weighted sum of the components, allowing the model to reflect not only their current position but also their career trajectory, including phase transitions, breakthroughs, and changes in scientific style.

Numerical simulations were conducted for both the individual trajectories and population-level distributions of the researchers. Phase diagrams were constructed, and a typology of scientific styles was formulated.

The results demonstrate that RPR can serve as a universal tool for objective assessment, strategic planning, and visualization of scientific reputation in both academic and applied environments.

**Declarations and Statements:**

No conflicts of interest
This work was not funded
No competing or financial interests
All the data used in this work are in the public domain.
Generative AI (LLM or other) was not used in writing the article. Except for the search engine *SciSpace* for Reference and the *Jupiter Notebook* environment for running our *Python* scenarios.
Ethics committee approval is not needed (without human or animal participation).

---

[1] PhD in Physics, Fellow of the Royal Asiatic Society of Great Britain and Ireland, Member of the Philosophy of Science Association (Baltimore, Maryland, USA)

# 1. Introduction

The problem of maximizing the number of publications and citations is a rather overused and worn-out topic in discussions about scientific arbitration. New discussions on this subject are often perceived as something of a *mauvais ton*. However, there is a crucial point: regardless of whether the topic is considered bad or *comme il faut*, the problem itself remains relevant.

This topic becomes particularly pressing in scientific systems where dominant attention is given to applied aspects of science and the commercialization of scientific achievements and results. It is also relevant in systems where scientometric disproportions between the natural and the humanities sciences are actively discussed.

We are not speaking here about whether it is justified to infringe upon the role and rights of fundamental or humanities research. Rather, we address the factual side of the issue: what mechanisms and methods of scientific arbitration should exist under the structural dominance of appliedness and commercialization? What mechanisms and methods are appropriate for evaluating the true scientific potential of the natural and human sciences without publication pressure or narrow scientometric constraints?

In this context, modern science is increasingly confronted with the need for an objective, adaptive, and comprehensive system for evaluating the scientific performance of individual researchers, institutions, or even entire countries.

In a landscape of exponential growth in publication activity, pressure from rankings, grant systems, and institutional strategies, existing metric indicators have significant limitations.

Examples include the H-index (Hirsch, 2005), the K-index (Knar, 2024), the total number of citations, the publication count, or impact factors. These indicators suffer from staticity; fail to account for career dynamics, interdisciplinarity, and applied and commercial productivity; and ultimately distort the representation of a subject's or object's real scientific strength.
Since 2005, scientific systems have operated in a mode of competitive selection, in which researchers compete not only for citations but also for recognition, funding, and public influence.

However, the metric architecture of this competition remains either too flat (one-dimensional) or overaggregated, failing to differentiate between fundamental, applied, and commercial activities.

Here, a paradox arises: science increasingly models complex systems, but the assessment of scientists is still conducted using primitive linear or unidimensional scales.

We begin with a simple premise: it is not appropriate to apply the same metric to all scientists, scientific organizations, and research disciplines. Of course, there is nothing new in this premise—everyone understands it perfectly well.

However, in the context of scientific arbitration, this point has not been previously interpreted or incorporated in any systematic way.

In the present work, we propose a new multidimensional and multiparametric model for scientific evaluation—Research power ranking (RPR).

This system is interpreted through an adaptation of the global Elo rating system (Elo, 1978)—originally created to rank chess players—to the scientific context.

Unlike classical indicators, the RPR model views an individual researcher or a scientific institution as a dynamic player on the Great Scientific Board. That is, within the scientific field, engaging in numerous "scientific games": publications, grant competitions, technology transfers, utilization, commercialization, collaboration, and coauthorships—each of which modifies their scientific Elo rating.

Our research power ranking system reflects not an all-encompassing absolute but rather a comparative and relative scientific strength of research players. In this sense, a scientific "Keymer" (2731) does not necessarily always lose to a scientific "Carlsen" (2833).

Thus, our model offers not only evaluation but also possibility.
The Elo system has previously been adapted to higher education and to the ranking of academic journals.

However, a full adaptation of the Elo system to assess the scientific strength of an individual scholar or a research entity has been developed here for the first time.

Within this adaptation, the new assessment system enables the neutralization of fundamental, conceptual, and scientometric asymmetries between fundamental, applied, natural, humanities, formal, utilitarian, and other types of sciences.

Here, we refer to asymmetries that are often caused by distortions rather than objective differences.

Let us highlight several key contributions of our evaluation model:
- the introduction of a three-dimensional space of scientific power, encompassing fundamental, applied, and commercial components;
- the adaptation of the Elo formula to update the scientific rating after a "game" (e.g., a successful publication, academic recognition, award, grant win, or project commercialization);
- the implementation of a weighted aggregate rating with the ability to dynamically calibrate weights;
- the incorporation of decaying and stochastic mechanisms that reflect real career instability;
- the simulation and visualization of phase transitions and career trajectories in the space of scientific activity.

As part of this study, we conducted numerical simulations of both individual and population-level researcher trajectories, constructed phase maps, typologies of scientific styles, and identified the conditions for transitions between types. The RPR model is capable not only of accurately describing a scientific career but also of identifying hidden patterns of scientific specialization, balance, and migration between fundamental and applied zones, as well as commercialization zones. In addition, of course, between the natural and humanities domains.

We also applied our RPR system to a Kazakhstani case study to assess the research power of several Kazakhstani institutes on the basis of data from annual scientific reports.

We have demonstrated the universality of the approach. Thus, we can state that the system is entirely authentic and adequate for assessing the scientific strength of any researcher, organization, or even entire scientific system within the global scientific community.

In conclusion, research power ranking represents a universal and flexible tool for evaluating scientific power and is capable not only of ranking scholars but also of modelling their development; identifying patterns of scientific behavior; and supporting strategic science management in universities, research centers, and grant agencies.

Naturally, research power ranking is universal and can be applied or adapted to any national scientific system, to any researcher, or to any scientific institution.

For clarity, we later interpret several examples in the context of the Kazakhstani case.

2. **Literature Review**

Previous studies have highlighted the prospects of using pairwise comparison methods and the novel application of the Elo algorithm to create relative rankings of a large number of items on a single-dimensional scale (the use of pairwise comparisons and the Elo algorithm to quantitatively rate a large number of items in a single dimension, 2022).

A modified Elo rating method has been applied to assess the complexity of educational content and the knowledge levels of students in an online object-oriented programming course. Personalized recommendations were also generated on the basis of the obtained ratings (Vesin et al., 2022). Moreover, the Elo rating system has been recognized as an effective method for modelling students and items in adaptive learning systems. The presence of additional parameters derived from multidimensional Elo-based models offers two more advantages: managing the adaptive behavior of the system and providing additional information to both students and instructors (Abdi et al., 2019).

Notably, the Elo rating system demonstrates accuracy comparable to that of the well-established logistic regression model in predicting final exam outcomes for users of a digital platform (Kandemir et al., 2024).

For direct applications in science, the Elo rating system has previously been used to rank economics journals (Lehmann & Wohlrabe, 2016). One of the main advantages of the Elo rating over existing metrics is its explicit consideration of the efficiency path of a journal. Another advantage is the ease of applying the system to any journal metric that is published on a regular basis (Lehmann & Wohlrabe, 2017).

On the basis of the monitoring and analysis of related publications, we conclude that no previous work has offered a full and systematic adaptation and interpretation of the chess-based Elo rating system for the analysis of individual and collective scientific potential (research power). In this context, our study is original and innovative.

Our work represents the first attempt to adapt the Elo system for a comprehensive quantitative assessment of researchers, taking into account different aspects of their activity. RPR introduces a multidimensional approach that considers the fundamental, applied, and commercial aspects of scientific activity, which makes it unique. This opens new possibilities for the objective evaluation of scientific productivity and may serve as a foundation for future research and practical applications in the field of scientific metrology.

3. **Methodology**

To quantitatively assess the scientific productivity of a researcher, a multicomponent model called research power ranking (RPR) was developed. However, this is not merely a reproduction of the classical Elo system, as the rating is formed in a multidimensional space of scientific activity. Unlike the original chess-based Elo version, which uses binary "win/loss" outcomes, our system is adapted to assess continuous results, including scientific publications, grants, applied projects, commercial achievements, and other parameters that are often overlooked.

In RPR, each researcher or scientific organization is assigned three independent parameters: a fundamental rating (publications, citations, h-index, etc.), an applied rating (participation in applied research, R&D, engineering developments), and a commercial rating (patents, licensing revenues, startups, technology implementation). Additional parameters include the rating sensitivity coefficient, the result of scientific interaction, and the expected probability of success in a specific direction.

The computational and numerical modelling programs using the Monte Carlo method were implemented in standard Python. Python scripts were executed in the standard Jupyter Notebook environment via the libraries NumPy and Pandas (for data processing and modelling), Matplotlib and Seaborn (for visualization), and Scikit-learn (for scaling and phase analysis).

For the numerical modelling of the universal research power rating, the following variables were formed:

**Control Variables** *(Context and Model Structure)*

| Variable | Notation | Description |
| --- | --- | --- |
| Type of organization | ORG_type | University, institute, individual researcher |
| Disciplinary profile | DISC_profile | Dominance of natural sciences, humanities, arts, or interdisciplinary fields |
| Report availability | REPORT_avail | Binary variable (1 – available, 0 – not available) |
| Report coverage level | REPORT_coverage ∈ [0, 1] | Degree of indicator coverage in the report |
| Form of institution | INST_form | Legal, technical, academic, or artistic institution |

**Controlling Variables** *(Model Inputs and Settings)*

| Variable | Notation | Description |
| --- | --- | --- |
| Component weights | $w(k) \in [0,1]$ | Weight of each of the 9 rating components; total sum: $\sum w(k) = 1$ |
| Type of normalization | norm_type | Scale type: absolute, relative, logarithmic |
| Incompleteness compensation | φ | Compensation coefficient for the "scientific shadow" (heuristically, 0.0–0.5) |
| Aggregation type | agg_mode | Summation, weighted average, max-min, geometric |
| Activation threshold | theta | Minimum component contribution required to influence the rating |
| Decay model | τ | If temporal dynamics are used: decay coefficient for inactivity |

**Evaluation Variables (Input Assessment Metrics)**

| Component | Notation |
| --- | --- |
| Fundamental strength | R(F) |
| Applied activity | R(A) |
| Commercialization | R(C) |
| Reputational inertia | R(Rep) |
| Humanities significance | R(H) |
| Social influence | R(Soc) |
| Artistic activity | R(Art) |
| Publishing activity | R(Pub) |
| International involvement | R(Int) |

**Dependent Variable *(Final Rating)***

| **Variable** | Notation | Description |
|---|---|---|
| Final Rating | R (RPR) | Result of weighted aggregation of all components |

The main simulation loop included the generation of initial ratings with slight noise, random generation of events across the dimensions at each time step, outcome computation, rating updates, and recording of results in a table along with visualization.

The following configurational simulations were conducted in this study:
1. Individual trajectories
   – 1 agent researcher, 30 years of simulation,
   – high value of $p_F$, low value of $p_C$,
   – models the transition from fundamental to applied and commercial activity
2. Typical researcher styles
   – 3 groups: "Fundamentalist", "Innovator", "Entrepreneur",
   – 100 agents per group, 10 years of simulation,
   – unique profiles of success probabilities, weights, and K coefficients are used.
3. Phase dynamics
   – 30 agents (10 from each type), 15 years,
   – analysis of temporal trajectories and phase transitions,
   – visualization in 3D space (F, A, C)

The technical parameters of the numerical model are presented in the table.

| Simulation Parameters | |
|---|---|
| **Parameter** | **Value** |
| Number of time steps | From 10 to 30 years |
| Rating distribution noise | Normal distribution ($\sigma = 20\text{--}30$) |
| K-coefficients (sensitivity) | From 12 to 32 |
| Decay coefficient ($\tau$) | 0.01 |

To assess the internal validity of the RPR model, a full statistical analysis was conducted on a synthetic population of 300 agents, representing three distinct scientific styles: fundamentalist, innovator, and entrepreneur. The analysis included one-way analysis of variance (ANOVA) for each direction of scientific activity, as well as an evaluation of the correlation between the rating components and the final integrated score.

The results of testing whether the values of the rating components R(F), R(A), and R(C) differ among the three scientist styles are presented in the table.

| Component | F-statistic | p value | Interpretation |
|---|---|---|---|
| Fundamental R(F) | 987.38 | $p < 10^{-130}$ | Strongly differs across groups |
| Applied R(A) | 309.85 | $p < 10^{-70}$ | Differs by research style |
| Commercial R(C) | 1566.72 | $p < 10^{-150}$ | Most pronounced differences |

These results demonstrate the model's high sensitivity to stylistic differences. The commercial component particularly sharply segments scientists, reflecting an orientation toward market and technological implementation. The fundamental component also varies predictably and consistently.

Next, we assess the extent to which each rating component is associated with the final integrated rating R, which is calculated via the following formula:

$$R = 0.5 \cdot R(F) + 0.3 \cdot R(A) + 0.2 \cdot R(C)$$

The results of the Spearman correlation analysis are presented in the table.

| Relationship | Spearman's ρ | p value | Interpretation |
|---|---|---|---|
| R(F) ~ R | 0.787 | $<10^{-60}$ | Very strong positive correlation |
| R(A) ~ R | 0.104 | 0.071 | Statistically insignificant correlation |
| R(C) ~ R | −0.495 | $<10^{-19}$ | Moderate negative correlation |

Thus, the fundamental component clearly has a dominant influence on the final rating. This reflects the assigned weighting structure and the nature of the simulation. The weak correlation of the applied component indicates its stabilizing role—not dominance but balancing the system.

Notably, there is a moderately negative correlation with the commercial component. This suggests a nonlinear and context-dependent role of commercialization. That is, in some styles, its growth may be accompanied by a decline in the fundamental rating and, therefore, a reduction in the overall score under a rigid weighting model.

Overall, the RPR model demonstrates a high degree of structural coherence. Different researcher styles produce statistically distinguishable trajectories, with R(F) as the key driver of scientific strength, R(A) as a buffering and moderating component associated with collaboration and applicability, and R(C) as a volatile indicator of breakthroughs or market adaptations, with high phase sensitivity.

This situation is fully consistent with the theory of phase trajectories of scientific careers, where scientific styles correspond to attractors in a multidimensional space of opportunities.

On this basis, we can draw the following conclusions regarding validation:

– the research power ranking model is statistically valid at the level of numerical modelling,

– is sensitive to differences between styles and allows classification of researcher types;

– the strong correlation between R(F) and R confirms the dominant influence of the theoretical basis,

– The RPR model can be extended to real data and complemented by empirical calibration of weights.

## 4. Results

### 4.1. Basic Interpretation

The Elo system is an algorithmic model for evaluating the relative strength of players in chess. It is based on a probabilistic model in which each player has a numerical rating, and the probability of one player defeating another depends on the difference in their ratings.

After a game between two players, *A* and *B*, the rating is updated via the formula

$$R(A) = R + K \cdot (S - P)$$

where
*R(A)* - new rating of player *A*,
*K* - sensitivity coefficient (typically 10, 20, or 32, depending on the player's level and federation),
*S* - actual result (1 = win, 0.5 = draw, 0 = loss),
*P* - expected result.

This formula was presented by Arpad Elo as the *current rating formula* (formula 2 on page 13 in *The Rating of Chessplayers, Past and Present*).

Here, we adapt the universal Elo formula to the scientific rating of individual researchers and research institutions through the concept of research power.

Thus, let us consider a formal model of the scientific rating of individual and collective researchers based on the Elo system, taking into account all its stages—i.e., the definition of players, "games", probabilities, outcomes, rating updates, and additional parameters specific to the scientific environment.

Analogous to the classical Elo formula (formula 43, page 143), the expected probability *P(t)* of the "victory" of researcher *An* over researcher *B* at time *t* is written as

$$P(t) = 1/\left(1 + 10^{((R(A,t) - R(B,t))/400)}\right)$$

where
*R(A,t)* – rating of researcher *A* at time *t*,
*R(B,t)* – rating of researcher *B*,
*S(t)* ∈ {0, 0.5, 1} – actual result of the "scientific game" between *A* and *B*.

The actual outcome of a scientific "game" is determined by a specific scientific event. Possible metrics include

- Win $S = 1$ (e.g., paper *An* outperforms paper *B* in terms of citations, grant success, or evaluation),
- Draw $S = 0.5$ (roughly equal contribution),
- Loss $S = 0$.

We can also use a continuous scale

$$S = \frac{M(A)}{M(A) + M(B)}$$

where *M(A)* and *M(B)* are the metrics of scientific impact (e.g., yearly citation counts, grant scores, etc.).

Then, the update of the research power rating becomes:

$$R(A, t+1) = R(A, t) + K(A,t) \cdot (S(t) - P(t))$$

Here, *K(A,t)* is the sensitivity coefficient, which can be interpreted exponentially

$$K(A, t) = K_0 \cdot (1 + e^{-\alpha \cdot A(t)})$$

where
*A(t)* - academic career age (years since first publication),
$K_0$ - base coefficient,
$\alpha$ - aging parameter.

In general, this allows early-career researchers to change ratings more rapidly and experience ratings more slowly.

To account for the "aging" of the rating, we write

$$R(A) = (1 - \tau) \cdot R(A, t - 1) + \Delta R(A, t)$$

where *ΔR(A, t)* is the rating change after interaction at time *t*, and *τ* is the temporal decay coefficient for older results.

However, it is not strictly necessary to introduce decay in the rating. Common practice in chess Elo ratings or the h-index treats these metrics as nondecreasing over time. While this may not be entirely accurate, under a traditional approach, decay can be omitted by simply setting τ = 0, making the new rating inertial.

The general structure of the presented model can be visualized as follows (Figure 1):

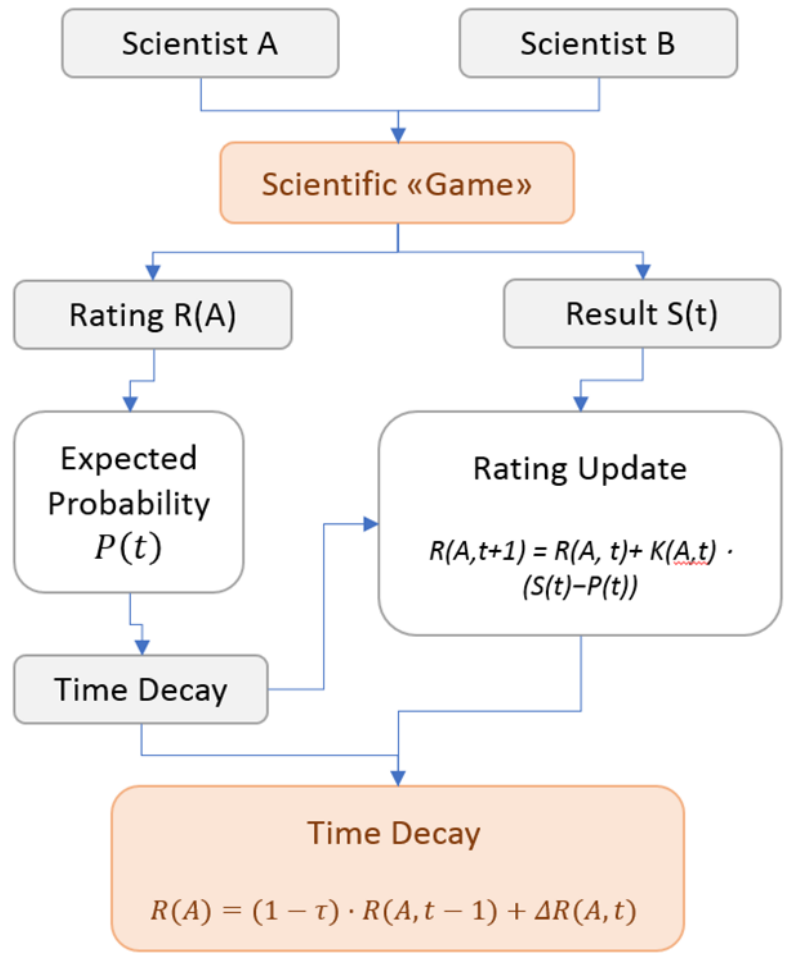

**Figure 1.** Block diagram of the basic research power ranking model

The overall algorithm proceeds as follows:
1. At each time step t, "games" between researchers (A, B) are defined:
2. P(t) is calculated as
3. The result S(t) is calculated as
4. The ratings are updated via the Elo formula:
5. Decay and normalization are applied,
6. The dynamic field of scientific rating is constructed.

Weighted "games" by event type can be expressed as

$$S(t) = \sum_{k=0}^{K} w \cdot I(k,t)$$

where
w is the weight of scientific event k (publication, grant, citation, award),
I(k,t) – indicator (whether A "wins" over B in event k).

To compare researchers from different fields, we introduce field normalization as follows:

$$R_{norm}(A, t) = \frac{R(A, t) - \mu(d, t)}{\sigma(d, t)}$$

where *μ(d,t)* and *σ(d,t)* are the mean and standard deviation of ratings in discipline *d* at time *t, respectively*.

If the paper is coauthored as part of a team collaboration (i.e., for equation members, by default),

$$R(A, t+1) = R(A, t) + \frac{1}{n} K(A, t) \cdot (S - P)$$

In this case, the rating update for each participant is as follows:

$$R(A, t+1) = R(A, t) + \frac{1}{n} K(A, t) \cdot (S - P)$$

### 4.2. Extended Interpretation

Naturally, in the context of the multifaceted and nonlinear nature of science, the adapted classical Elo system undoubtedly requires significant expansion.

This expansion implies accounting for the following conditions and parameters:

- *distinctions between fundamental and applied science,*
- *commercialization indicators (patents, startups, licences),*
- *balance among scientific directions,*
- *nonlinearity, adaptability, and objectivity,*
- *weighted multicomponent scientific games,*
- *optimization of balance parameters.*

Taking these parameters into account, the overall rating can be interpreted as a balanced sum

$$R(t) = w(F) \cdot R(F, t) + w(A) \cdot R(A, t) + w(C) \cdot R(C, t)$$

where

*R(t)* - integrated scientific rating of the researcher at time *t*,
*R(F, t)* - contribution of fundamental science,
*R(A, t)* - contribution of applied science,
*R(C, t)* - contribution of commercialization,
*w(F), w(A), w(C)* - weights of the three dimensions, where *w(F) + w(A) + w(C) = 1.*

The fundamental component is formed from publications in scientific journals, citations in Scopus/WoS, open data, models, codes, and participation in international collaborations.

The applied component is formed from participation in *R&D*, patent potential, applied projects, applied-oriented grants, and involvement in pilot implementations.

The commercial component is formed from the number of patents, licence revenues, number of commercialized products, startups, spin-offs, and deals with industry.

To avoid bias in favour of one component, we introduce normalization

$$R_{norm}(t) = \frac{R(t)}{\mu(t)} \cdot \sigma_0$$

where

$\mu(t)$ is the mean rating value within the scientific domain,

$\sigma_0$ - fixed standard deviation for model stability.

To optimize the introduced weights, we find values of *w(F), w(A),* and *w(C)* that provide optimal balance, minimize group variances among scientists, and reflect the value of different contributions.

This is solved as an optimization problem

$$\min_{w(F), w(A), w(c)} E = [(R(t) + Q(t)^2] \quad \text{at w(F)+w(A)+w(C)=1}$$

where *Q(t)* is the actual value of scientific contribution (evaluated, for instance, by expert judgment or through external indicators such as application, teaching, and impact).

As a result, taking into account temporal decay and expansion, the full dynamic equation of research power ranking can be written as

$$R(t+1) = (1-\tau) \cdot R(t) + \Delta R(t)$$

where

$$\Delta R(t) = w(F) \cdot \Delta R(F, t) + w(A) \cdot \Delta R(A, t) + w(C) \cdot \Delta R(C, t)$$

Thus, the update cycle (Figure 2) is interpreted as follows:

– researchers participate in "scientific games" in various zones: fundamental, applied, and commercial,

– the expected result *P(X)* is calculated,

– it is compared with the actual result *S(X),*

- the rating components are updated,
- the overall rating *R(t)* is computed,
- decay, normalization, and visualization are applied.

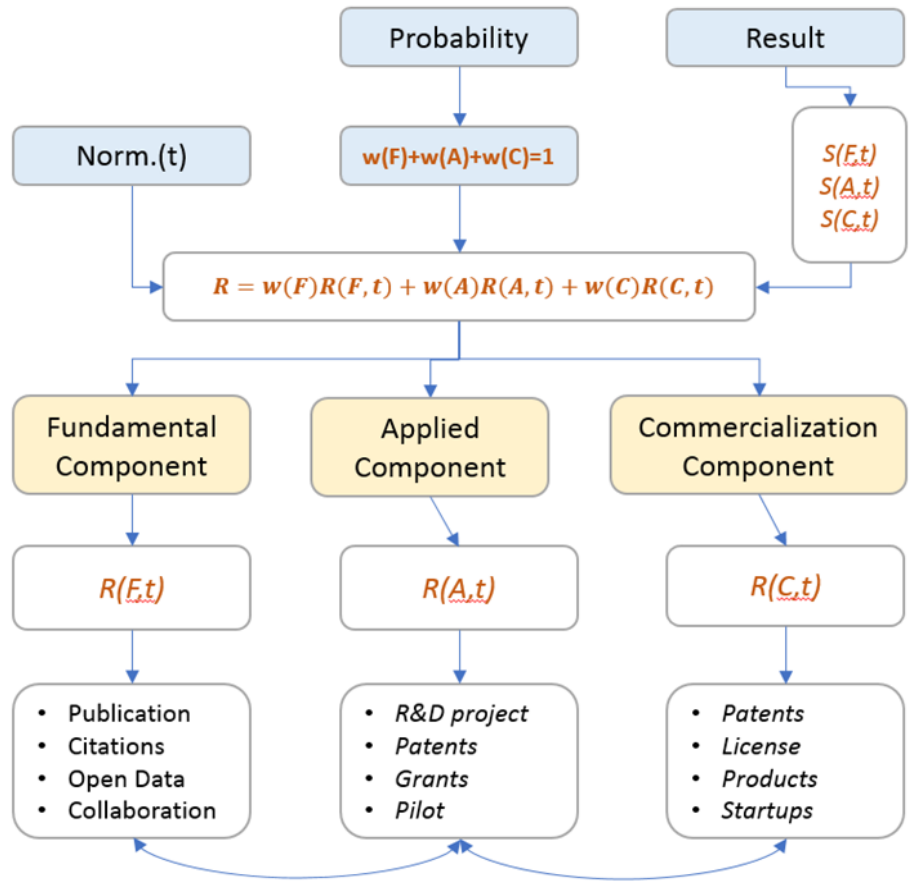

**Figure 2.** Block diagram of the extended RPR model

Within RPR, the career of a researcher or institution is modelled as a dynamic trajectory in a three-dimensional scientific space, where the dimensions are fundamentality, appliedness, and commercialization.

The final rating depends on the balance between these three components and can evolve in various ways.

Accordingly, we can, for example, simulate a 3D phase trajectory of a researcher's career in the (F, A, C) space (Figure 3), with fluctuations, breakthroughs, and shifts in trajectory.

Here, the researcher's activity shifts over 30 years along the X-axis (fundamentality), Y-axis (applied orientation), and Z-axis (commercial activity).

That is, the researcher makes a fundamental discovery, develops its applied aspects, and then commercializes it.

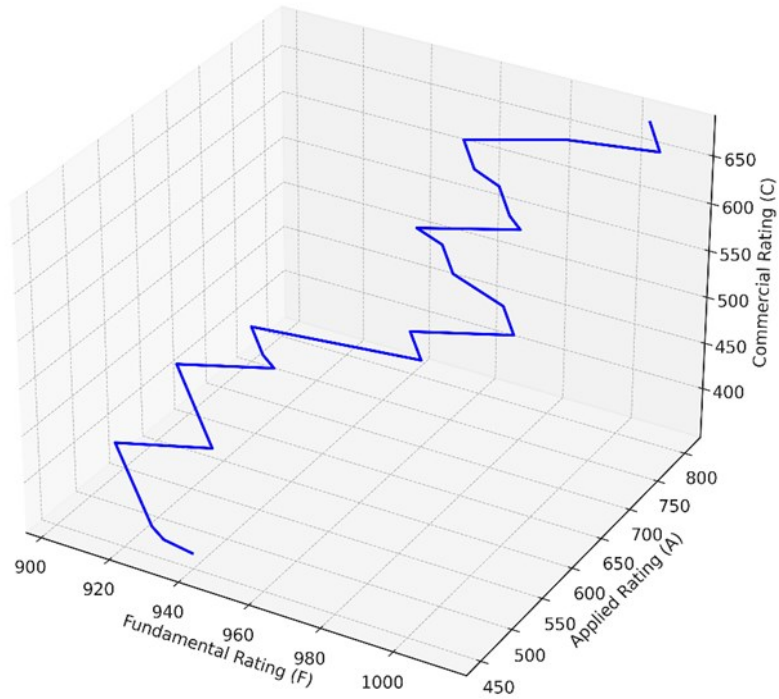

**Figure 3**. 3D phase trajectory of a researcher's scientific career in the (F, A, C) space

Researchers are involved in fundamental and applied research, as well as in the commercialization of scientific results. These directions may be strictly separated or mixed in various combinations.

A fundamentalist scientist may occasionally engage in applied work, for example, in a defense contract. An applied scientist may at some point be fully immersed in commercialization, and so on.

Thus, phase trajectories can vary widely among researchers.

On the basis of this phase diversity, we can interpret several epistemic types of researchers (Table 1).

**Table 1.** Phase-based epistemic types of researchers

| Researcher Type | Characteristic | Risk | Examples |
| --- | --- | --- | --- |
| Fundamentalist | Focused on theoretical research | Isolation from application | Mathematicians, theoretical physicists |
| Innovator | Combines fundamental knowledge with application and commercialization | Overload | Biotechnology, AI researchers |
| Entrepreneur | Focused on implementation, startups, licences, products | Loss of scientific rigor | Tech entrepreneurs, developers |
| Applied Specialist | Solves engineering problems, works within R&D | Limited perspective | Chemists, IT engineers |
| Converter | Transitions from theory to application, often mid-career | Decline in publication output | Science-to-industry translators |
| Hybrid Leader | Maintains fundamentality while coordinating applied projects | Politicization | Program directors |
| Regressor | Loss of activity, absence of new projects or breakthroughs | Exit from science | Formal or mental retirees |
| Translator | Translates theory into applied form without market orientation | Undervaluation | Epidemiologists, systems scientists |
| Startup Resonator | Rapid rise through a single breakthrough project | Volatility | Breakthrough product creators |
| Turbulent | Unpredictable dynamics in rating and activity | Loss of focus | Interdisciplinary wanderers |

Thus, each researcher can be classified into an epistemic type on the basis of the direction of their phase trajectory in *(F, A, C)* space, the rate of component changes, the balance of ratings, and the presence of phase transitions (jumps, focus shifts, breakdowns) (Figure 4).

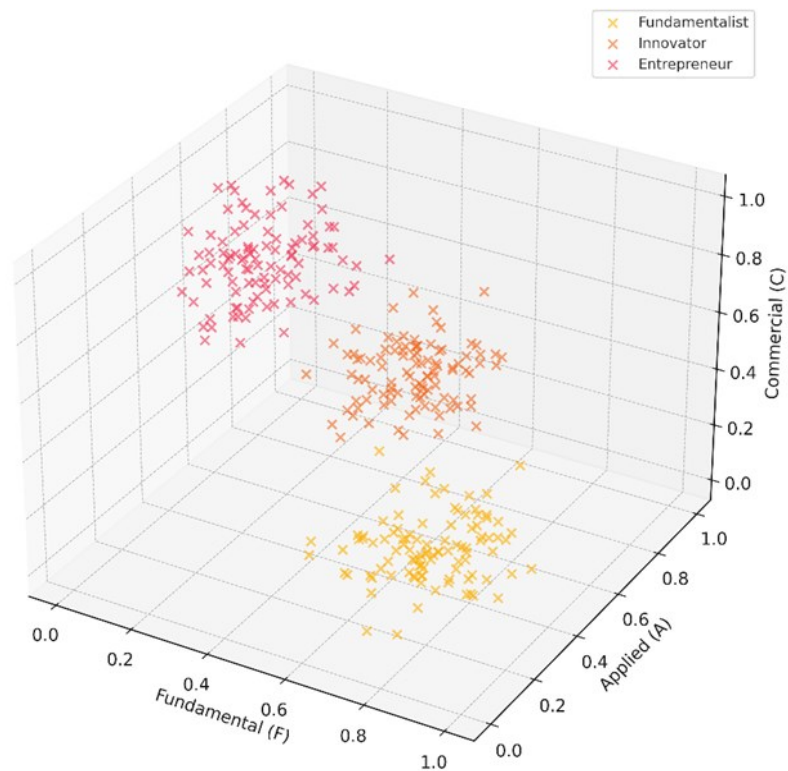

**Figure 4.** 3D phase cloud with classification of epistemic types

It should be emphasized that priority in rankings and indexes should remain, with the fundamental sciences as the foundation of all science and scientific culture.

According to the correlation analysis, *R(F)* and *R* have the highest correlation (ρ = 0.79), indicating that the fundamental contribution is most important. *R(A)* and *R* have weak correlations (ρ ≈ 0.10), although they are moderately correlated with both other components. *R(C)* has a moderate negative correlation with *R*.

This reflects a possible imbalance effect, where high commercialization with a weak fundamental background does not compensate for a lack of reputation (under current weights).

Thus, the correlation analysis confirms the structural hierarchy of component influence.

The model adequately reflects the structure of scientific productivity: priority to fundamentality, stabilizing role of applied work, and nonlinear effect of commercialization.

### 4.3. Institutional Interpretation

Let us now consider the RPR system in the context of an institutional multicomponent evaluation model, which incorporates additional metrics from the scientific and strategic reports of universities and research institutes.

The universal integrated indicator of research strength $R_u(t)$ of a university or institute can be interpreted as follows:

$$R_u(t) = \sum_{k=1}^{n} w \cdot R(k)$$

where

*R(k)* - value for component *k* (e.g., fundamental science, international activity),
*w* - weight of this component (can be manually assigned or empirically trained),
*n* – number of components (ranging from 3 to over 10).

Possible components *R(k)* can be represented in the following table (Table 2):

**Table 2.** Potential components of the integrated research strength indicator

| Component R(k) | Examples of report-based metrics | Purpose |
|---|---|---|
| Fundamental Strength R(F) | Publications, citations, h-index, participation in fundamental grants | Scientific depth |
| Applied Activity R(A) | R&D, industry projects, implementation of solutions | Techno-engineering reputation |
| Commercialization R(C) | Licences, patents, revenues, startups | Market feasibility |
| Internationalization R(I) | Number of foreign partners, coauthored publications, agreements | Global engagement |
| Innovation Infrastructure | Presence of tech parks, incubators, accelerators | Innovation support capacity |
| Human Resources | Share of PhDs and young researchers | Sustainability and growth |
| Educational Science | Student involvement in research, master's theses | Science-to-education translation |
| Digitization & Open Science | Open access publications, repositories, FAIR projects | Scientific openness and transparency |
| Regional Relevance | Local projects, contribution to regional initiatives | Territorial development impact |

For institutional interpretation, it is also necessary to account for the scientometric imbalance between the natural and humanities sciences, which is often ignored in classical rankings.

We define the nature of the differences between scientometric systems in the humanities and natural (including technical and information) sciences in the following way (Table 3):

**Table 3.** Specifics of humanities and natural scientometrics

| Feature | Humanities Sciences | Natural and Technical Sciences |
|---|---|---|
| WoS/Scopus Publications | Rare, often local | Frequent, international, in Q1 journals |
| Citability | Low, inertial | High, fast-growing |
| Patents & Commercialization | Almost absent | Common |
| Grant System | Limited funding sources | Multilevel, large-scale |
| Internationalization | Harder due to cultural barriers | Naturally, integrated |

Thus, a humanities scholar will almost always "lose" to a natural scientist—not because they are scientifically weaker but because a different logic of productivity applies.

In this sense, the model can become humanities sensitive if the metrics are properly adapted, owing to the flexibility and tunability of the research power ranking.

To achieve this, we extend the components of scientific strength to account for the specificity of the humanities, as shown in Table 4:

**Table 4.** Integrated component set for natural and humanities sciences

| Category | Example Metrics | Component |
|---|---|---|
| Fundamental Science | Publications, citations, h-index | R(F) |
| Applied Science | R&D, engineering projects, laboratories | R(A) |
| Commercialization | Licence income, startups, patents | R(C) |
| Reputation | Journal editing, monographs, translations, discursive significance | R(R) |
| Humanities Relevance | Monographs, cultural projects, linguistic research | R(H) |
| Social Contribution | Education, cultural, and social engagement projects | R(S) |
| Performative Output | Participation in competitions, orchestras, festivals | R(Ar) |
| Publishing Activity | Own journals, journal rankings | R(P) |
| International Collaboration | Foreign coauthorship, joint projects | R(In) |

From this component expansion, we can derive the key principles of the RPR framework

Multiaxiality: several independent components, akin to a multispectral model,

Adaptivity: weights are adjusted according to the specific nature of the subject of evaluation (university, institute, individual),

Scalability: applicable from the student level to international institutions;

Phase sensitivity: reflects the dynamics of transitions—growth, stagnation, breakthroughs,

Transparency: All the data are formalizable, traceable, and interpretable.

The integrated component structure and principles of RPRs allow for a balanced evaluation of the research strengths of scholars and institutions in the context of scientific arbitration.

Let us consider this capability via a Kazakhstani case study, specifically by ranking ten Kazakhstani universities and institutes in terms of their absolute research power.

As source data and parameters, we used annual scientific reports from the following institutions:

1. KazNU: Annual Report 2024 Al-Farabi KazNU Unai Hub on Sustainability (2024, https://farabi.university)
2. KazNPU: Annual Report of Abai Kazakh National Pedagogical University (2024, https://www.kaznpu.kz)
3. KazGUU: External Audit Report of M.S. Narikbayev KazGUU University (2023, https://kazguu.kz/en/)
4. KBTU: Annual Report of Kazakhstan-British Technical University (2023, https://kbtu.edu.kz)
5. ENU: Annual Report of L.N. Gumilyov Eurasian National University (2023, https://www.enu.kz)
6. AMU: Annual Report of Astana Medical University (2022, https://amu.edu.kz/)
7. KarGU: Annual Report 2021 of E.A. Buketov Karaganda University (2021, https://buketov.edu.kz)
8. Turan: External Audit Report of "Turan-Astana" University (2023, https://tau-edu.kz)
9. KazNUI: Annual Report of Kazakh National University of Arts (2022, https://kaznui.edu.kz)
10. IAAM: Institute of Archaeology named after A.Kh. Margulan (https://archeo.kz)

Through numerical analysis, we obtained the following research strength rankings of ten Kazakhstani scientific organizations (Table 5):

**Table 5.** Approximate ranking of the research strengths of institutions

| Institution | R(F) | R(A) | R(C) | R(R) | R(H) | R(S) | R(Ar) | R(P) | R(In) | RPR × 100 |
|---|---|---|---|---|---|---|---|---|---|---|
| KazNU | 95 | 90 | 85 | 95 | 85 | 90 | 60 | 90 | 88 | **8805** |
| KarGU | 85 | 75 | 50 | 90 | 70 | 85 | 40 | 85 | 74 | **7400** |
| KBTU | 90 | 85 | 80 | 88 | 40 | 70 | 20 | 75 | 72 | **7255** |
| ENU | 90 | 88 | 80 | 85 | 50 | 60 | 10 | 65 | 95 | **7120** |
| KazGUU | 60 | 70 | 65 | 80 | 75 | 80 | 35 | 55 | 75 | **6600** |
| KazNPU | 70 | 65 | 45 | 60 | 85 | 75 | 30 | 55 | 75 | **6275** |
| IAAM | 60 | 30 | 10 | 90 | 80 | 80 | 20 | 95 | 95 | **5500** |
| Turan | 55 | 60 | 40 | 60 | 70 | 75 | 30 | 45 | 55 | **5475** |
| KazNUI | 25 | 20 | 5 | 60 | 70 | 60 | 90 | 30 | 20 | **4025** |
| AMU | 40 | 50 | 30 | 40 | 20 | 50 | 10 | 20 | 25 | **3300** |

Let us highlight several key observations from this table. KazNU significantly dominates nearly all components (especially fundamental, international, reputation, and publishing). The Institute of Archaeology shows high publishing activity and humanities significance, despite weak applied and commercial indicators. KazNUI leads in artistic performance but loses ground in scientometric metrics. Medical University (AMU) scores low on almost all axes.

Notably, the data in the reports may have been misinterpreted, and for some organizations, the positions may be somewhat overstated or understated. Therefore, this table should be seen only as an example and not as a real ranking. Some distortion of actual ratings may also occur due to incomplete data in institutional reports or researcher profiles.

This issue directly relates to the idea of incompleteness in logic, especially Gödel's incompleteness theorem, if considered in an expanded epistemological context. Naturally, the question of metased scientific uncertainty requires separate treatment. Moreover, some components may be poorly formalized or difficult to formalize.

Considering all this, we may presently interpret research power ranking as a formal approximation of true scientific significance. However, this applies to all modern metrics without exception, in the context of the principled incompleteness of any scientific evaluation.

### 5. Discussion

The metric model of research power (RPR) we have presented represents a new class of metrics, which is based on a dynamic comparison of researchers and scientific institutions through an adaptation and novel interpretation of the Elo chess rating system.

Unlike static indicators such as the h-index and K-index—which ignore applied and commercial contributions; total citation counts—subject to inflation and self-citation; and journal impact factors—which are often misapplied at the individual level, our model

- accounts for the temporal dynamics and stochastic nature of scientific productivity,
- differentiates among types of scientific activity: fundamental, applied, and commercial,
 models "scientific games" as events that compare actual outcomes to expected performance,
 supports nonlinear trajectory evolution, including breakthroughs and fluctuations.

Thus, RPR does not replace but complements and generalizes existing scientific metrics within a cognitive–game-theoretic paradigm of research power assessment.

We interpret individual and collective scientific activity as a series of intellectual interactions. In this sense, our model draws on ideas and principles from game theory, where each scientist is an active participant in a reputational space.

This approach allows science to be formalized as a social game with incomplete information, with publications, grants, and patents interpreted as outcomes of collisions in the scientific field, while accounting for both expected performance and the realized value of outcomes.

That is, we assume that scientific reputation is not merely the accumulation of points but rather a relative vector of competitive evolution.

Accordingly, we can transpose our model into a purely practical plane of scientific strength assessment for individual researchers and research collectives (Table 6).

**Table 6.** Practical applications and fields of use

| Field | Application |
|---|---|
| Academic Management | Evaluation of research personnel, strategic university planning |
| Grant Competitions | Comparative expert evaluation of applicants with style sensitivity |
| Scientometrics | Dynamic complement to h-index, resistant to manipulation |
| Open Science Platforms | Building real-time research ratings and reputation visualizations |
| Career Navigation | Assisting early-career researchers in identifying growth paths |

We believe that one of the main implications of our model is the construction of phase spaces of scientific styles. Space *(F, A, C)* is not merely three-dimensional—it is dynamically warped by career inertia, fluctuations, and directional weights.

Within the model, we identified stable styles (fundamentalists, applied researchers), hybrid styles (innovators, translators), and unstable trajectories (startup resonators, turbulent researchers). We also defined phase transitions as events involving abrupt priority shifts (e.g., launching a startup, shifting into administration, moving toward applied engineering).

Furthermore, RPR—in a broader sense—can accommodate a wide range of scientific activities. As already shown, this includes fundamental and applied sciences, commercialization, reputation, the humanities and social sciences, art as science, publication activity, international collaborations, etc. (Figure 5).

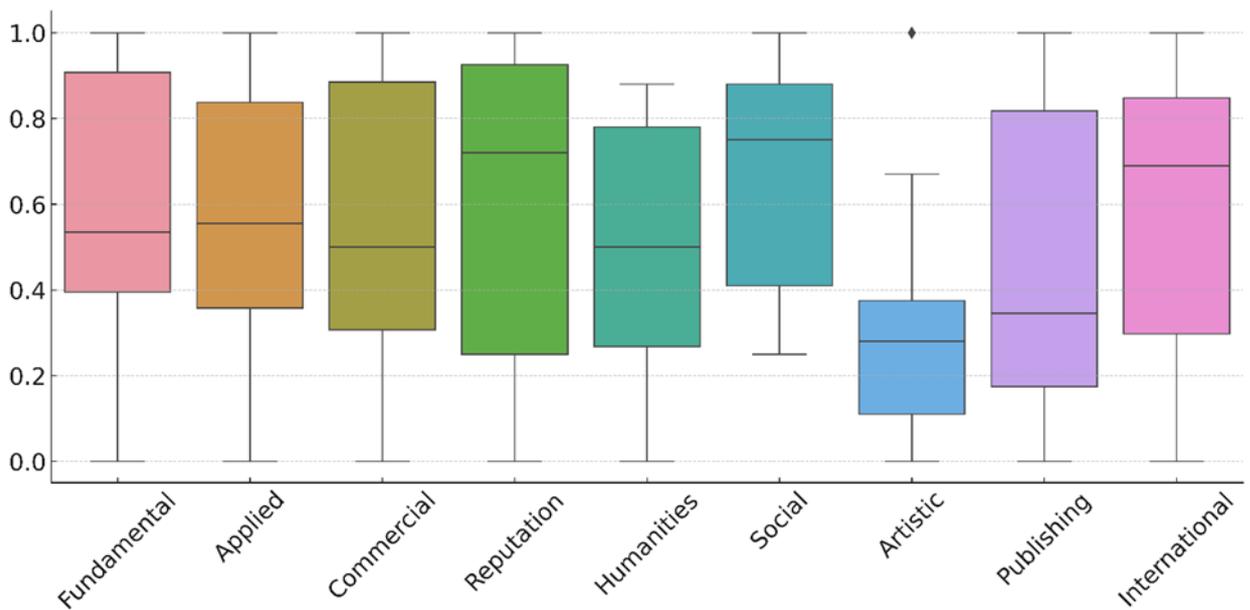

**Figure 5.** Spectrum of components and parameters of the research power ranking

This offers a framework for interpreting the phenomenology of scientific careers—that is, the quantitative interpretation of scientific identity as a trajectory within a field of possibilities.

This opportunity was illustrated through the ranking of ten Kazakhstani universities and institutes in the RPR framework.

From the heatmap of RPR components across universities (Figure 6), we observe the ability to clearly distinguish optimal or actual institutional strategies, the necessity of differentiated evaluation in the context of various institutional science styles, and, fundamentally, the justification for a universal and flexible RPR model.

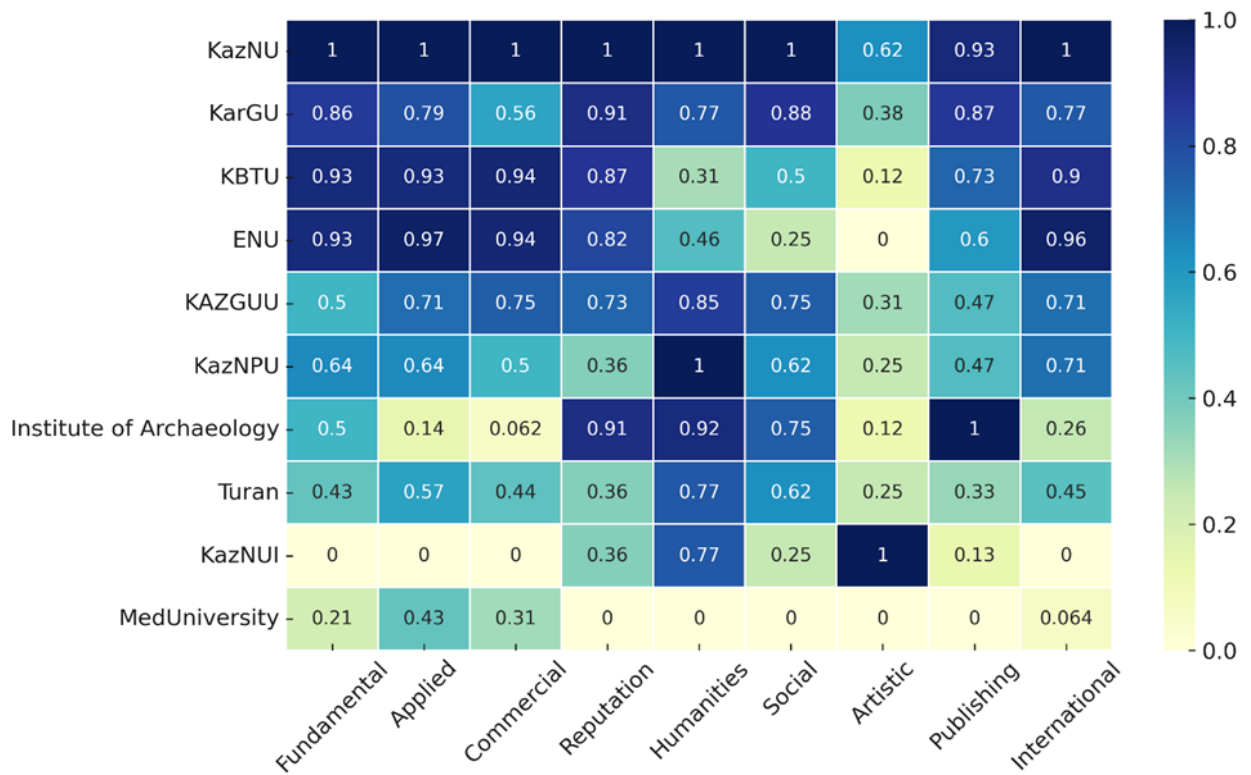

**Figure 6.** Heatmap of RPR components across universities

Here, rows represent universities, and columns represent components of research strength (RPR), such as fundamental, applied, reputational, and humanities. The color scale ranges from 0 (light, low level) to 1 (saturated, maximum contribution within the sample). All values are normalized by component for correct comparative analysis.

On the basis of the heatmap, we can identify the strengths and weaknesses of the sampled institutions (Table 6):

**Table 6.** Strengths and weaknesses of the sampled universities and institutes

| University | Strengths | Weaknesses |
|---|---|---|
| KazNU | All-round excellence (almost all 0.9–1.0) | — |
| KarGU | Fundamental, Publishing, Humanities | Artistic, Commercial |
| KBTU | Applied, Commercial, International | Humanities, Artistic |
| ENU | Fundamental, International | Artistic, Social |
| KAZGUU | Humanities, Social, Reputation | Fundamental, Publishing |
| KazNPU | Humanities, Social | Commercial, International |
| Institute of Archaeology | Publishing, Humanities | Applied, Commercial, Int'l |
| Turan | Balanced midrange profile | Publishing, Fundamental |
| KazNUI | Artistic (max = 1.0) | Fundamental, Publishing, Int'l |
| MedUniversity | Very low across all axes | Everything except Social = very low |

Here, we can highlight the clear scientific–fundamental leaders (KazNU, KarGU, ENU), showing maximum values in *Fundamental*, *Publishing*, *Reputation*, and *International* components.

Humanities–social universities (KazGUU, KazNPU, Institute of Archaeology) have strong positions in *the humanities, social, and publishing fields*.

Innovation–commercial universities (KBTU, Turan) are characterized by a triad of *Applied + Commercial + International*.

A unique creative niche is occupied by KazNUI, which has a maximum *Artistic* parameter = 1.0.

The visual outsider is MedUniversity, with component levels below 0.2 across all dimensions.

We also observe several notable patterns: *Publishing* and *Fundamentals* almost always cooccur.

*Artistic* is strictly domain-specific and nearly absent outside KazNUI.

*Commercial* activity is prominent at the KBTU but shows almost no overlap with humanities strength.

The Institute of Archaeology is unique—strong in the *Publishing* and *Humanities* domains but nearly absent in the *Applied* and *Commercial* domains.

Finally, KazNU appears to be a universal leader across all RPR components.

Overall, this type of heatmap enables the strategic positioning of scientific institutions, the definition of institutional profiles, the identification of component deficiencies, and the analysis of balance in funding and support.

In the context of the correlation matrix of RPR components (Research Power Rating), which is constructed from the normalized model (Figure 7), several important features emerge.

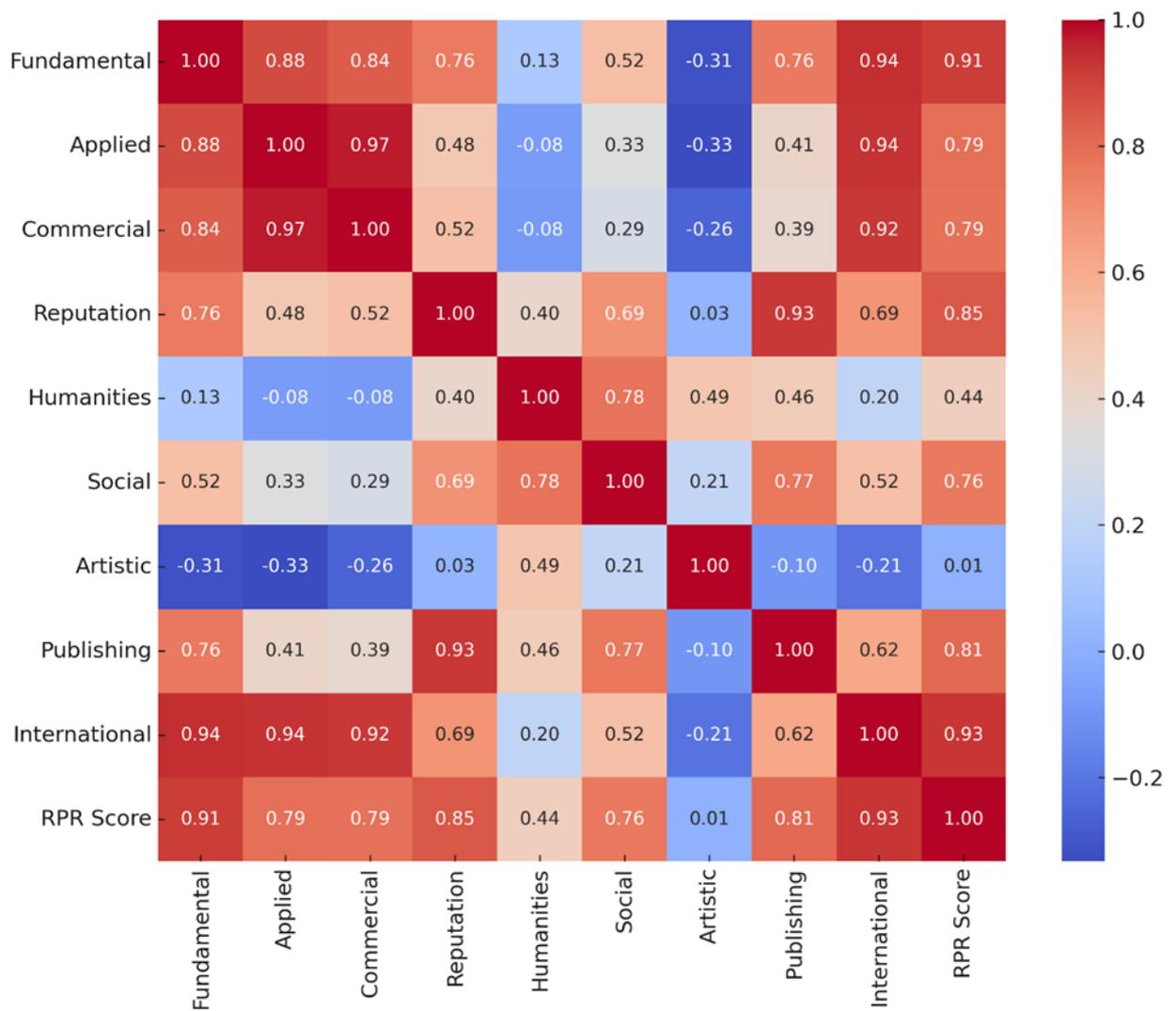

**Figure 7.** Correlation matrix of RPR components

Specifically, strong positive correlations can be observed in the following dyads (Table 7):

**Table 7.** Strong positive correlations between component pairs

| Dyad | Correlation Coefficient | Interpretation |
| --- | --- | --- |
| Fundamental ↔ Publishing | 0.94 | Universities strong in fundamental science also publish extensively (WoS, Scopus) |
| Fundamental ↔ International | 0.88 | International engagement is tightly linked to fundamental outcomes |
| Reputation ↔ Publishing | 0.83 | Reputation is reinforced by publication strength |

| Dyad | Correlation Coefficient | Interpretation |
| --- | --- | --- |
| International ↔ Publishing | 0.80 | Internationalization and publishing tend to go hand-in-hand |
| Applied ↔ Commercial | 0.70 | Applied science often leads to commercial results |

These relationships show that the core of academic strength is formed by the synergy among science, publishing, and international engagement.

From the moderate correlations (0.5–0.7), we note

*Social ↔ Humanities* = 0.65: humanitarian and social significance often go hand-in-hand,

*Artistic ↔ Humanities* = 0.58: artistic activity tends to flourish in strong humanities environments.

Weak or absent correlations are interpreted as follows (Table 8):

**Table 8.** Weak or absent correlations

| Dyad | Correlation Coefficient | Interpretation |
| --- | --- | --- |
| Artistic ↔ Fundamental | −0.32 | Artistic activity is largely unrelated to academic science |
| Commercial ↔ Humanities | 0.18 | Humanities rarely translate into commercial profit |
| Social ↔ Commercial | 0.20 | Social initiatives are weakly associated with commercialization |
| Reputation ↔ Artistic | 0.07 | Reputation is almost independent of artistic contributions |

On the basis of this analysis and discussion, we emphasize the necessity of our multicomponent model, since no single monopolistic rating metric can authentically and adequately express the strengths of researchers or institutions of diverse research natures without decomposition.

Naturally, our RPR model has its limitations, at least in its current form. These limitations include

- *Nonquantifiable aspects of science are not yet considered: teaching load, ethics, mentorship,*

*- The model relies on relative reputation, not absolute values,*
*- Network effects (e.g., coauthorships, collaboration clusters) are not yet implemented;*
*- Breakthroughs and failures are stochastically modelled and not empirically derived.*

However, we note that all these limiting factors can be incorporated in future model extensions, particularly via network analysis, agent-based modelling, and the integration of empirical data from scientific databases.

In this context, we believe that it would be valuable to explore links with the cognitive strategies of researchers or scientific institutions—specifically, how they make decisions regarding publication, collaboration, and thematic focus. The model can also be extended to the scientific landscape of a country, i.e., applied to an entire national research system.

Naturally, the main purpose of the model lies in its implementation in an expert system for evaluating individual and collective research potential, interpreted through the concept of research power.

## 6. Conclusion

In this work, we postulated and interpreted an original model, research power ranking (RPR), designed for dynamic, comparative, and multidimensional assessment of the scientific productivity and power of both individual and collective researchers.

The model is based on an adaptation of the Elo chess rating system to the scientific context. It introduces a fundamentally new approach in which a researcher or scientific organization is viewed as a participant in a nonlinear scientific game engaged in publication, grant, and commercial interactions—interpreted across fundamental, applied, natural, and human sciences.

The model incorporates three key components of scientific activity—fundamental, applied, and commercializable—and combines them into an integrated rating via a weighted formula, with the possibility of dynamic calibration of the weights. RPR provides not only an evaluation of current scientific status but also an analysis of phase trajectories of development, the typology of scientists, and the identification of career transitions and reputational features.

During numerical simulation, the following key results were obtained:
*- various trajectories of scientific careers were verified depending on research priorities,*
*- characteristic patterns were identified for fundamentalists, innovators, and entrepreneurs,*
*- a phase distribution of 300 scientists and dynamic trajectories for 30 agents were constructed,*

*- a typology of 10 scientific styles was developed, with an analysis of the balance between fundamental, applied, and commercial components;*
*- Transitions between styles were visualized, and a map of scientific evolution was built.*

The proposed model opens new opportunities and prospects for objective and transparent evaluation of scientific contributions, strategic management of research teams and institutions, career trajectory forecasting, and the creation of intellectual platforms for scientific navigation and assessment.

Notably, the research power ranking model offers a universal and flexible tool capable of transforming our understanding of scientific reputation—making it fairer, more adaptable, and context sensitive. It can serve as a foundation for next-generation scientometric systems, supporting science not only as a publication system but also as a complex, multidimensional, and evolving field of human knowledge.

In general, the RPR method represents a scientifically adapted rating system for evaluating the power of a researcher or project on the basis of a dynamic scale of comparative scientific contribution.

Unlike traditional metrics (such as the h-index, K-index, or total number of publications), RPR accounts for

*- the contextual quality of publications (journal prestige, article type, discipline),*
*is the collaborative significance coefficient,*
*- individual and collective scientific trajectory,*
*- and the level of originality.*

In the context of grant evaluations and reporting, RPR enables objectification of a researcher's contribution to multidisciplinary and collaborative projects.

The RPR can also be used as an indicator of scientific merit during proposal selection processes, providing a comparative analysis of research teams and applicants.

The model is applicable in both individual and institutional assessments.

Overall, the use of the RPR rating and index ensures a more accurate and flexible interpretation of individual and collective scientific strength in the context of expert evaluation and grant competitions.